\documentclass[a4paper,11pt]{article}
\usepackage{pos}

\title{A generalized approach to study low as well as high $p_T$ regime of transverse momentum spectra}
 \ShortTitle{Application of Pearson Distribution in high energy collision}

\author*[a]{Rohit Gupta}
\author[a]{Satyajit Jena}

\affiliation[a]{Indian Institute of Science Education and Research\\
  Mohali, Punjab, India}

\emailAdd{ph15049@iisermohali.ac.in}
\emailAdd{sjena@iisermohali.ac.in}

\abstract{A good understanding of the transverse momentum $(p_T)$ spectra is pivotal in the study of QCD matter created during the heavy-ion collision. Considering the difference in the underlying particle production mechanism, $p_T$ spectra can be divided into two distinct regions. Low-$p_T$ region corresponds to particle produced in soft-processes whereas particles produced in hard processes dominate the high-$p_T$ regime of the spectra. We will discuss a unified formalism to explain both low as well as high-$p_T$ region of the transverse momentum spectra in a consistent manner. This unified formalism is based on the generalisation of non-extensive statistical mechanics using the Pearson distribution. This generalised formalism also gives a strong insight into the study of elliptic flow in heavy-ion collision. Content of this proceeding is primarily based on the work \cite{Jena:2020wno}.}

\FullConference{%
  40th International Conference on High Energy physics - ICHEP2020\\
  July 28 - August 6, 2020\\
  Prague, Czech Republic (virtual meeting)
}

\begin{document}
\maketitle
\section{Introduction}
Quark Gluon Plasma (QGP) is defined as a state where quarks and gluons are free to move inside a nuclear volume rather than only in nucleonic volume. The transition from hadronic state to QGP occur at the phase boundary where the critical temperature $(T_c)$ is sufficient enough to support the formation of QGP droplet. Search for $T_c$ and the type of phase transition is being explored in the experiments by scanning the QCD phase diagram. \\
In order to study the QCD phase diagram, temperature of the system is required, which can be extracted from the transverse momentum spectra $(p_T)$ of final state particles. Hence an appropriate theoretical description of $p_T$ spectra is of prime importance among particle physics community. Quantum Chromodynamics (QCD) is the underlying theory to explain the strongly interaction QGP system, however, due to the asymptotic freedom and the very nature of QCD coupling constant, the coupling strength is very strong at low-$p_T$ values and hence we cannot apply the perturbative theories to study the $p_T$-spectra in this regime. Hence, we resort to the phenomenological approach with most accepted being the statistical thermal models. 
\section{Theoretical description of $p_T$-spectra}
Considering the system produced in high energy collision is of purely thermal origin, most natural choice is to apply standard Boltzmann-Gibbs (BG) statistics. Distribution function of transverse momentum spectra in BG statistics is given as:
\begin{equation}
\frac{1}{2\pi p_T} \frac{d^2 N}{dp_T dy} = \frac{gV m_T}{(2 \pi)^3}  exp\left(- \frac{m_T}{T}\right)  
\label{eq:bg}
\end{equation}
It has been observed that the data of $p_T$ spectra deviate heavily from the function Eq. (\ref{eq:bg}) at low as well as high $p_T$. Also, BG statistics is only applicable in the system where the number of particles is of the order of Avogadro number which is not the case in heavy-ion collision where number of final state particles is of the order of few thousands only. \\
A generalisation of Boltzmann statistics to include the non-extensive system was put forward by Tsallis \cite{Tsallis:1987eu}. Tsallis formalism includes an additional parameter $`q'$ which takes care of non-extensivity in the system. This additional parameter also acts as a scaling factor to make standard statistical mechanics applicable on system with low number of particles. Although, the Tsallis distribution function for $p_T$ spectra (Eq.\ref{eq:tsa})
\begin{equation}
\frac{1}{2\pi p_T} \frac{d^2 N}{dp_T dy} = \frac{gV m_T}{(2 \pi)^3} \left[1+(q-1)\frac{m_T}{T}\right]^{-\frac{q}{q-1}} 
\label{eq:tsa}
\end{equation}
 fits the data in a consistent manner as compared to BG distribution, it is applicable only upto a certain range of $p_T$ corresponding to particles produced in soft-processes. On the other hand, we have a well established perturbative QCD based power-law form of the distribution function to explain the spectra in high $p_T$ regime which is dominated by the particles produced in hard scattering processes. \\
 We have utilized the Pearson statistical framework \cite{Pearson343} to propose a generalization of Tsallis distribution to explain both soft and hard part of the spectra. Pearson distribution is a master equation given in the form of a differential equation.
Solution of the Pearson differential equation has been modified and the corresponding distribution function for $p_T$ spectra is given as: 
\begin{equation}
 \label{eq:pearson_final}
\frac{1}{2\pi p_T} \frac{d^2 N}{dp_T dy} =  B \left( 1 + \frac{p_T}{p_0}\right)^{-n} \left( 1 + (q-1)\frac{p_T}{T}\right)^{-\frac{q}{q-1}}
\end{equation}
This generalized formalism is thermodynamically consistent and backward compatible. Here, backward compatibility means that the generalized formalism reduces to Tsallis formalism under some limit on parameters. \\
\section{Result and Conclusion}
\begin{figure}[h!]
 \centering
  \begin{minipage}[b]{0.4\textwidth}
    \includegraphics[width=\textwidth]{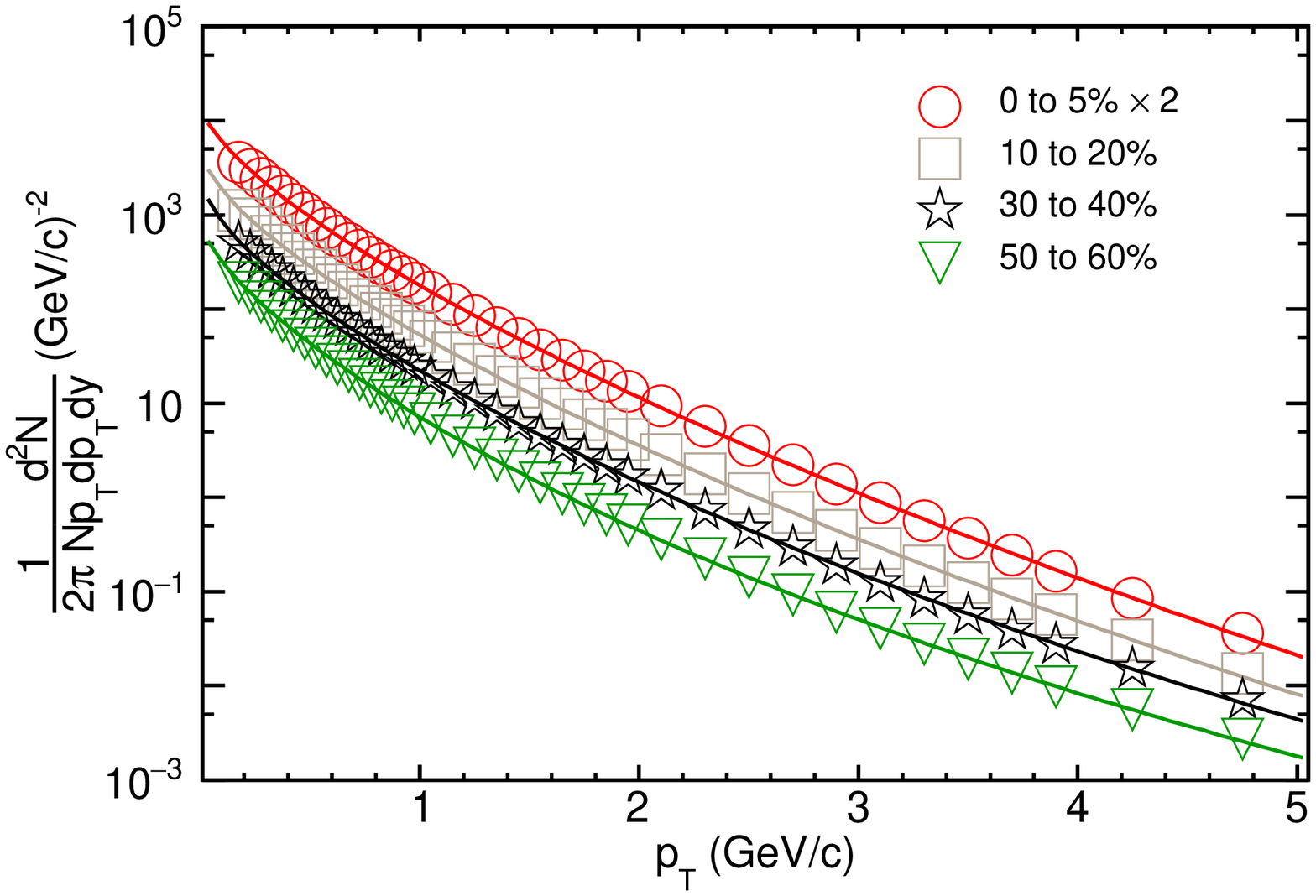}
    \caption{Pearson Fit to $p_T$-spectra of charged hadrons produced in $2.76$ TeV Pb-Pb collision \cite{Abelev:2012hxa} at four different centrality.}
    \label{fig:pearson}
  \end{minipage}
  \hfill
  \begin{minipage}[b]{0.4\textwidth}
    \includegraphics[width=\textwidth]{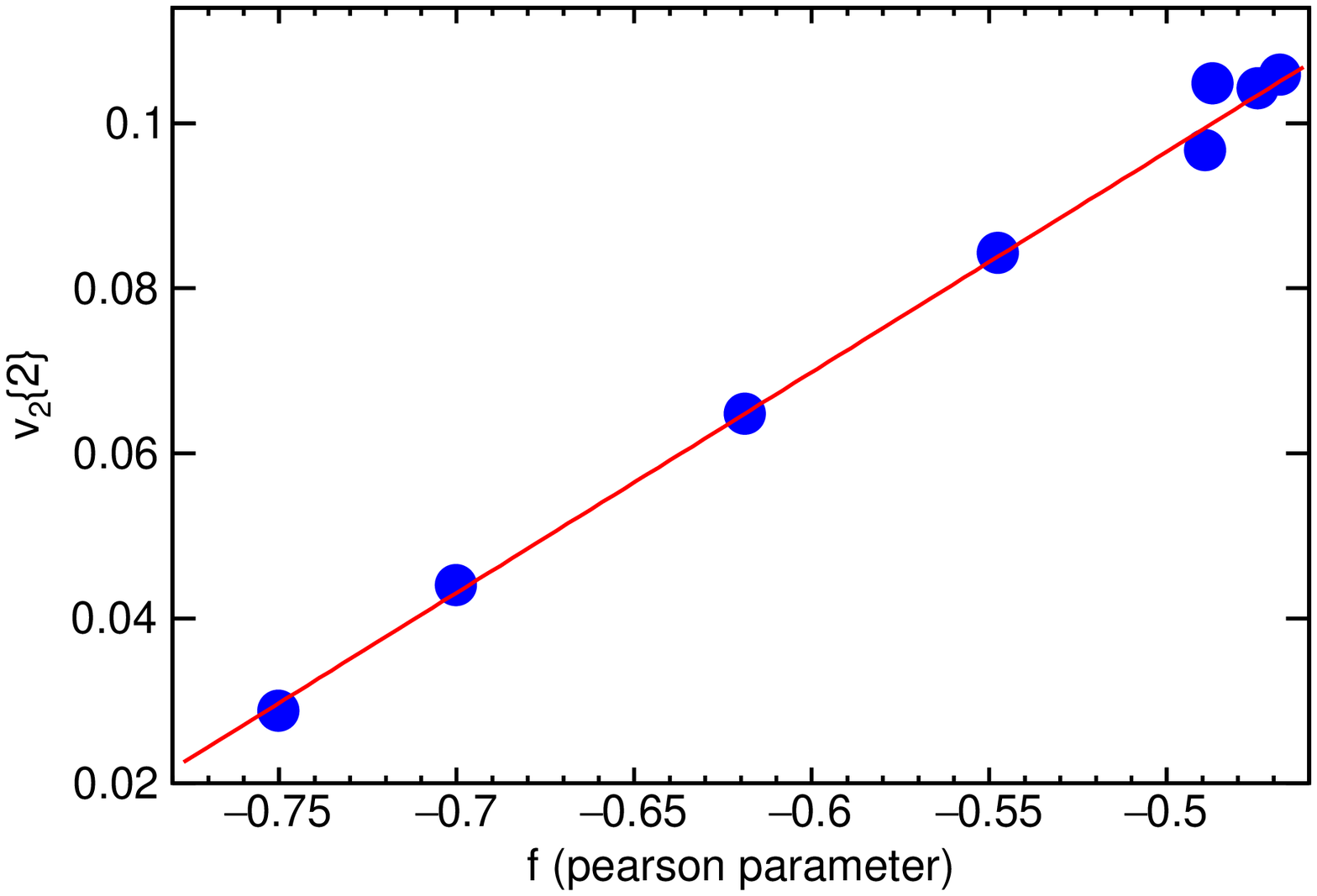}
    \caption{Elliptic flow coefficient versus Pearson parameter f at 2.76TeV Pb-Pb collision and the curve is fitted with a linear equation.}
    \label{fig:flow}
  \end{minipage}
  \end{figure}

The fit of $p_T$ spectra with Eq. (\ref{eq:pearson_final}) has been presented in Fig. (\ref{fig:pearson}) and we observe that the generalized distribution nicely fit the spectra over a wider range of $p_T$ as compared to Tsallis distribution.
  
 We also observe that one of the parameters in generalized distribution is linearly related to the elliptic flow coefficient as presented in Fig. (\ref{fig:flow}).\\
 To conclude, this formalism has a unique potential to explain both hard as well as soft part of $p_T$ spectra in a unified manner without violating the basic thermodynamics principles. 

\end{document}